# #BlackLivesMatter and Racism in Life Expectancy, Poverty, Educational Attainment, and Race Compositions: State Analysis of 2020 Tweets in the USA

## Author Information

Kalpdrum Passi, Associate Professor, Laurentian University, kpassi@laurentian.ca , ORCID ID 0000-0002-7155-7901

Shervin Assari , Center for Research on Ethnicity, Culture and Health, School of Public Health, University of Michigan, Ann Arbor, MI, USA,  assari@umich.edu , ORCID ID 0000-0002-5054-6250

Amir Hossein Zolfaghari, Laurentian University, azolfaghari@laurentian.ca , ORCID ID 0000-0003-2913-1330

## Abstract

The year 2020 was a challenging year known mainly as the pandemic year. However, the notable event of George Floyd's killing broke many humans' hearts and made them protest on social media and the streets as well. In this research, we studied the hashtag "BlackLivesMatter," and some of its adversary contentions regarding George Floyd's demise in 2020 on Twitter. Based on the extensive aftermath of protests in the United States, we considered an area analysis to compare tweet rates in different groups to some previously studied statistics. The purpose is to investigate how racism content is correlated with life expectancy, poverty, and education. Findings revealed a significant relationship between online color-based contents and some physical world indicators.

# Introduction

The story of #BlackLivesMatter gets back to an evening of February 2012, when Trayvon Martin [6] decided to tour around, and a neighbor who was suspiciously watching him ended up shooting this innocent Black teenager. The fact that he was unarmed became highlighted, and as that killer was found unguilty in the summer of 2013, some people protest against this justice failure. That made Alicia Garza, Patrisse Cullors, and Opal Tometi construct #BlackLivesMatter and popularize it on social media - a call to fight for Black people's rights [1]. However, the same story reoccurred repeatedly, such as the killing of Michael Brown [2] by a police officer in Fergusen, Missouri, in August 2014, or a more recent one George Floyd [3], who was killed on May 25, 2020.

#BlackLivesMatter is the flag of those demanding a change in the world against racism and is a figure in social movements for fighting against racial injustice.

## Why Study on Racism?

Humans need to express opinions, feelings of pride and pain, and intellectual powers. That is how we join a community of people to have a sense of value and influence. In this day and age, this demand emerges in a microblogging service that allows users to declare their views with short messages that roughly resemble their ideas.

Twitter provides a free, fast, and federal messaging service that allows worldwide to share their 280 characters note in parallel with other 335 million active users [4]. This is the path that population uses to express themselves that sometimes can show the undercover beliefs.

But what are the consequences? There is an increasing number of researches that show inequalities leading to health problems. The situation is the same even in the well-developed states of Europe, and the resolution of socioeconomic health disparities is a significant cause of public health disappointments. [5].

Researchers identified a term for this purpose: "Social Gradient [6]" which defines a condition when individuals face less socioeconomic situations and worsen their health than those who are more advantaged. Eventually, it shortens their lives and poses them with severe health risks. It is currently among the primary robust research outcomes in public health, as manifestations confirm health inequalities enduring during life and across societies lead to health problems [7].

Many studies [8–13] have shown that social networking sites are powerful platforms to research health either as a medium for health care interventions, users' behavior studies, or health content analysis. Twitter is the favorite platform of at least 22% of Americans [14]. Regarding American adults, 61% of them are using the virtual world as a source of health information [15],[16], whereas Twitter is the most commonly used social media in this specific topic [17].

This study tried to spotlight the connection of racism-content tweets and four already available tables from previous research. This study aimed to see the relation between content published on Twitter and some real-life factors with links to the life of Black people.

## Literature Review

**Racial content on Twitter**

Criss et al. [18] studied the effects of racial content on Twitter with their participants and showed the adverse effects on mental, emotional, and physical health. They stated that even some Twitter users with racial interests tend to cooperate to create an "echo chamber" to reinforce racial content. A striking result is the feelings of disappointment, rage, and sorrow in other users, which compromise their psychological health, comprising depressive indications, tensions, and nervousness. Ferrara et al. [19] support this view, pointing out content published on social media has emotional influences on the audiences. They observed random Twitter users and found that emotional contagion aroused negative and positive sentiments in other users.

Williams and Mohammed, [19] in 2009, attempted to see the evidence and research gaps about discrimination and health. They found that racism can affect people's health negatively facing discrimination. Discrimination and health have an opposite association after reviewing the available articles on racism and health indexed in PubMed between 2005 and 2007.

Racial health disparities are widespread; referring to the Kung et al. study (2008) [21], Black people in the US are suffering more death rates among principal causes of death than whites. Considering the fifteen foremost reasons that take human life, such as cardiovascular conditions, cancer, stroke, diabetes, renal failure, hypertension, liver cirrhosis, and homicide, the rates are disappointingly high. African Americans and American Indians have higher death rates by age than whites. Likewise, in 2001, Levine et al. [20] found that almost 100,000 black people die prematurely each year because of racial health disparities. Although life expectancy for blacks and whites has increased, the gap between the races in life expectancy is high, about 5.1 years in 2005, according to the National Center for Health Statistics, 2007.

Similarly, the infant death rate has declined for blacks and whites, but the racial gap has increased since 1950 [21]. Williams believes that the results of researches about health issues show that the disparities are even getting worse. As reported by NCHS [21], statistics show that a clear example is heart disease and cancer, which are the two significant causes of death in the United States. Evidence indicates that black and white people had different death rates in these circumstances in 1950; however, now, African Americans have even more death rates than whites.

Joseph-Shehu et al. [12] explored the effects of the information and communication technologies tools such as social networks, websites, text messages, and cell-phone apps on health-promoting lifestyle behavior and reported its positive impact. They determined the essential impression of these technologies on preventing and managing diseases, healthy BMI, and its rule to promote physical and psychic health.

Scientists found the application of Twitter research beneficial for medicine because it enables medical professionals to reach a broad audience, such as physicians, trainees, or patients [22,23]. Anyone with an account can access broad terms of information, reply, ask questions, add favorites, or even retweet them to share them with friends. These features can address two advantages, direct medical professionals to the content with the most reaction and guide users to reliable sources.

Still more remarkable, we have "hashtags" indicating the tweet conversation's topic category, which encourages other users to continue discussing that disease or question type. The potential of Twitter to share and advance biomedical research also made several medical experts employ it for its features that allow them to promote their research, new treatments, or clinical difficulties that require further study, panel discussion, or investigations [17].

## Methodology and Analysis

We conducted an ecological data analysis study utilizing Twitter data in the first six months of 2020 to compare with national "Life Expectancy" as well as "Poverty Rate," "Educational Attainment," and "Race Composition" tables. We aimed to see how these tweets categorized as racism or related to the BLM movement are associated with these tables. We gathered these tweets using an inclusion criterion and following enhanced features utilizing geography and gender database APIs. We prepared and cleaned the data through different steps, and finally, we calculated the rates for the states and compared them using Spearman Rho correlation analysis by IBM SPSS Statistics. We studied the 2020 tweets, a time before and after the death of George Floyd [24], and we used an archive tool that is a 1% sample from the Twitter stream API daily tweets.

Generating a dataset to study social media starts with defining inclusion criteria. We represented three groups regarding some specific words included in the tweet's text or appended as hashtags. We were considering only tweets that had geographical data inside the United States and dated in 2020. We fetched the records based on three main groups "BlackLivesMatter (BLM) Movement," "Anti-BLM Movement," "Racist," and an extra one that is the ambiguous group, while each of these has its search terms.

First and foremost, we have our "BlackLivesMatter" or "**BLM**" in short, group, which are those who stand in line with this movement by their hashtags. Hence, we included all tweets with this hashtag or have it inside in content. Facing this movement, some other users reacted by opposite hashtags "AllLivesMatter," "WhiteLivesMatter," or "BlueLivesMatter." We consider them the second group of the study named "**Anti-BLM**." Some minorities discussed this event by a tweet that employs hashtags from both groups, categorized as "**Ambiguous**."

Finally, we have the "**Racism**" group covering those tweets that used a series of old, ugly, offensive words known as n-words in their tweets [25]. These n-words can carry significant psychological effects and provoke severe harm on the audience [26]. However, regarding the writing notation or pronunciation, there are multiple forms of these n-words that we referred to the dictionary enlisting [27] to shape our criteria and incorporate them in the study.

Next, we used the Python Geopy library [28] to determine the associated state with the users' tweet profile locations across the US. We also obtained genders by using their username and employing Gender-API [29], which includes 4 million names, and achieved the highest performance compared to its contenders [30]. The frequency of available tweets for each state was calculated. We also measured the total number of published tweets for each state to calculate the rates.

Further, we analyzed the rates from our study and compared them with four tables mentioned before, using the Spearman Rho correlation test that is a nonparametric test for calculating the correlation. Spearman correlation analysis was performed between this study's outcome rates based on states, and "Life Expectancy," "Poverty Rates," 'Educational Attainment," and "Race Compositions," tables while a 2-sided $\alpha < 0.05$ was considered statistically significant. A two-tailed test and consideration of 0.05 as the significance level means that this test would share half of this significance level (0.025) to test the statistical analysis of one direction and the other half for testing the significance in the different direction [31]. The resulting tables are marked with ** for those results that could have a significant correlation at the 0.01 level (2-tailed) and * marks, denoting the significant findings at the 0.05 level (2-tailed).

**Life Expectancy**

The "life expectancy" term addresses the number of years that a newborn expects to be alive. While this rate cannot be identified at birth, researchers determine an average lifetime of years for members of a specific population to acquire this number. This factor is an essential and prominent measure in research and policy that is widely used to indicate health status [32], [33]. The results were compared with "life expectancy at birth" data [34] employing the "National Center for Health Statistics" (NCHS) records.

**Poverty Rate**

Many black Americans have a sad story at birth, experiencing an environment where poverty is typical, educational failure dominates, and social deterioration overflows. Continuous exposure to such an environment diminishes the chance of a Black fellow's social or economic success story [35]. We employed a list of U.S. states and the associated poverty rate [36] to compare and see its correlation with our dataset. We applied three available lists of poverty rates in the U.S., which are:

1. 2018 Poverty rate, percent of persons in poverty [37]
2. 2014 Poverty Rates that include unrelated children [37]
3. Supplemental Poverty Measure, an average measure of 2010–2014 with Geographically Adjusted [38]

**Educational Attainment**

Statisticians commonly use the term "educational attainment" to indicate an individual's highest grade of education. It is used chiefly as an average equivalent for a group or territory. The corresponding information is accomplished by a simple question that asks, "What is the highest grade of school you have completed?" and we derived these statistics using a survey done by the American Community Survey [39] from 2013 to 2017 [40].

**Race Composition**

We used statistics from "World Population Review" data [41] and specifically compared the White and Black population rates in each state to the tweets we acquired. These tables are statistics regarding the composition percentage of Black and White people in each state.

## Results

The inclusion criteria yielded 306,925 tweets out of the 43,830,301 tweets with location values available inside the United States of America in the first six months of 2020 using archived data. The four data groups acknowledged before and after the George Floyd killing event. Considering gender, different rates were obtained for each state and the total tweets published in the associated time.

The most extensive data group is the Racist with 229,924 records, and the following ones are BLM, Anti-BLM, and Ambiguous with 69,116; 7,617; and 267 records, respectively (Table 1). Gender specification was achieved for 52.31% of the data, which is 160,539 records.

.

*Table 1. Number of tweets in each group detailed by month*

| Month | Black Lives Matter | Anti BLM | Ambiguous | Racist | Tweet Numbers | Sum of all GeoTagged Tweets |
|---|---|---|---|---|---|---|
| January | 202 | 73 | 1 | 43950 | 44226 | 7,544,965 |
| February | 96 | 112 | 2 | 20740 | 20950 | 3,604,863 |
| March | 117 | 115 | 0 | 42638 | 42870 | 8,401,976 |
| April | 250 | 218 | 0 | 42245 | 42713 | 8,194,289 |
| May Before 25 | 340 | 528 | 1 | 31554 | 32423 | 6,056,215 |
| May After 25 | 15957 | 1588 | 186 | 9838 | 27569 | 1,985,889 |
| June | 52154 | 4983 | 77 | 38959 | 96173 | 8,042,104 |
| **Sum** | **69,116** | **7,617** | **267** | **229,924** | **306,925** | **43,830,301** |

The first visible outcomes are the impressive number of tweets with the #BlackLivesMatter after George Floyd's demise on May 25, 2020 (Figure *1*). The numbers substantially rose to 68,111 in 35 days compared to 1,005 tweets in the first 145 days of the year. Considering that this hashtag movement's history gets back to 2013, this study's outcomes show a significant increase in tweets published after this sad event.

Similarly, those who stand on the other side with their Anti-BLM tweets were more active after this event. Although this group number is small, it jumped to 6,571 from 1,046, comparing the one-month and four-day period after the event to the four months and 25 days before that. On the other hand, the Racist group had a slight decrease compared to the other groups. Taken all groups together regarding the event's impact, results are compared to show the trends.

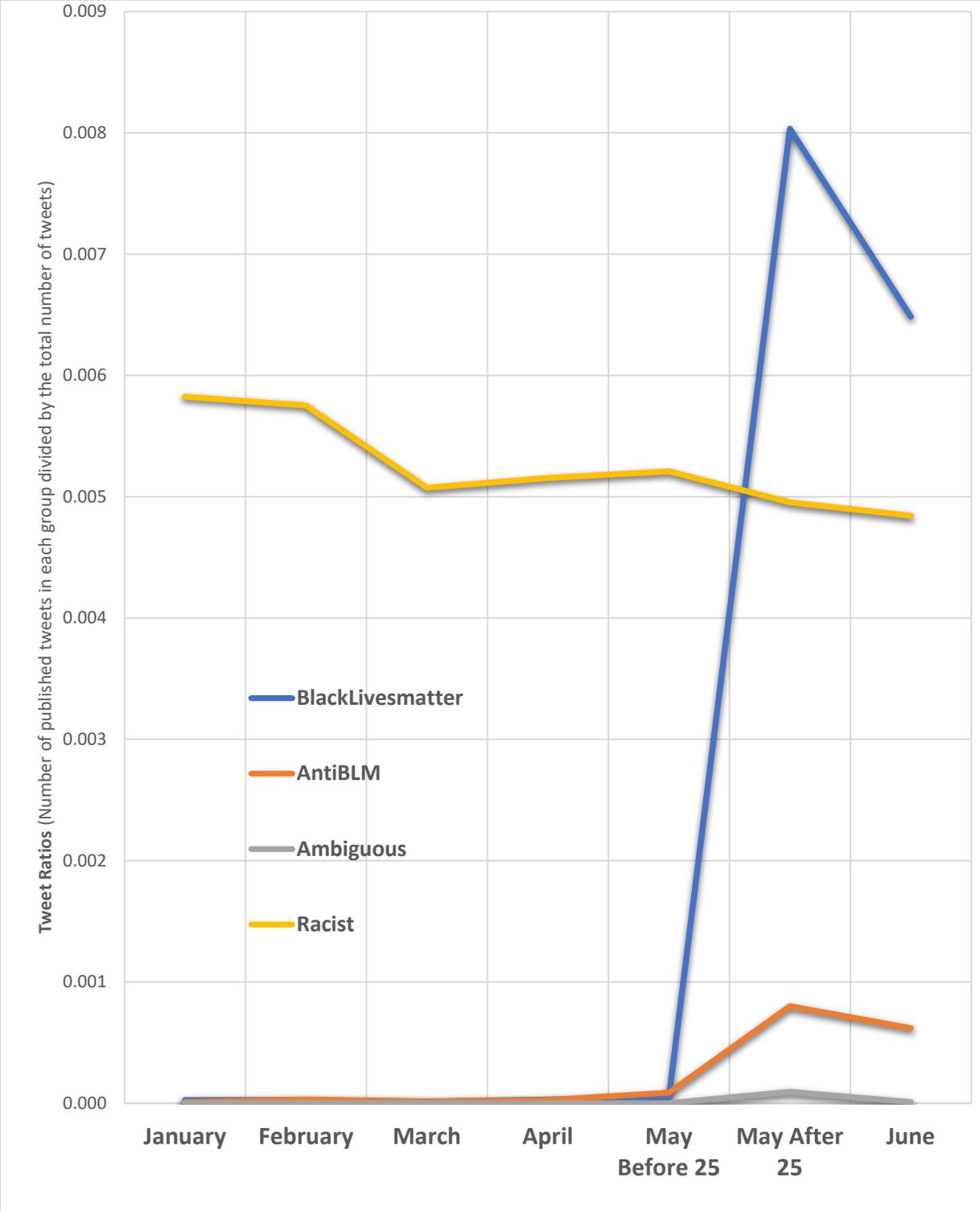

Figure 1. Compares the trends of each group in 2020

Subsequently, the outcomes of the study were analyzed to national "Life Expectancy," "Poverty Rate," "Educational Attainment," and "Race Composition" statistics. There is strong evidence of a relation between racist content rates and the life expectancy of male black people. The more surprising correlation is also with poverty rates, high school educational assessment, and race composition. The single most striking observation from the data comparison was a significant positive correlation between the number of bachelor's or advanced degree people and the number of tweets in the BLM group. The higher the degree level of an area, the more users participated in this movement defending blacks.

**Correlation with Life Expectancy**

The outcomes show a positive correlation between the racist tweets and "Calculated Life Expectancy of White Male – Black Male." From this correlation, we can understand that there is a trend between these factors, and as when the racist tweet numbers are more extensive, there is an enormous difference between White and Black's life expectancy. For instance, the result is significant at the $p < 0.01$ level by the Spearman correlation coefficient of 0.408 for 40 states with the available data for Black life expectancy (Table 2).

Moreover, the rate of BLM tweets has a positive correlation with Black life expectancy. It alternatively implies that in areas where people tend to publish more BLM tweets and care more about this movement, Black people have higher life expectancy rates. The rate of BLM tweets is determined by dividing the sum of all BLM tweets by the sum of all tweets published over the study time. This rate has a significant correlation at the $p < 0.05$ level by the Spearman correlation coefficient of .357 with "Black expectation of life at age 0" and at P-value $< 0.05$ level by the coefficient of .347 with "Black male expectation of life at age 0" (Table 3).

Table 2. Spearman's rho statistical analysis for percentage of racist tweets compared to life expectations of "White people minus Black" regarding gender.

|  |  | Number of Racist tweets Male | Number of Racist tweets Female | Racist tweets Male Before event | Racist tweets Female Before event | Racist tweets Male After event | Racist tweets Female After event | Racist rates (Racist / Sum) |
|---|---|---|---|---|---|---|---|---|
| Calculated life exp. Of Whites – Blacks | Correlation Coefficient | 0.156 | -0.064 | 0.162 | 0.177 | -0.046 | 0.158 | 0.219 |
|  | Sig. (2-tailed) | 0.336 | 0.695 | 0.318 | 0.276 | 0.778 | 0.329 | 0.174 |
|  | N | 40 | 40 | 40 | 40 | 40 | 40 | 40 |
| Calculated life exp. Of White Male – Black Male | Correlation Coefficient | .412** | -0.181 | .416** | .435** | -0.162 | .418** | .408** |
|  | Sig. (2-tailed) | 0.008 | 0.265 | 0.008 | 0.005 | 0.317 | 0.007 | 0.009 |
|  | N | 40 | 40 | 40 | 40 | 40 | 40 | 40 |
| Calculated life exp. Of White Female – Black Female | Correlation Coefficient | -0.223 | 0.227 | -0.217 | -0.202 | 0.225 | -0.231 | -0.093 |
|  | Sig. (2-tailed) | 0.166 | 0.160 | 0.178 | 0.211 | 0.163 | 0.152 | 0.567 |
|  | N | 40 | 40 | 40 | 40 | 40 | 40 | 40 |

Table 3. Spearman's rho statistical analysis for rate of BLM and Anti-BLM tweets regarding gender, compare to "life expectations," "Poverty Rates," "Educational Attainments," and "Race Composition" statistics.

|  |  | Rate of all BLM tweets (BLM / Sum) | Rate of BLM Male | Rate of BLM Female | Rate of Anti-BLM Male | Rate of Anti-BLM Female |
|---|---|---|---|---|---|---|
| Black expectation of life at age 0 | Correlation Coefficient | .357* | .366* | 0.287 | 0.245 | 0.107 |
|  | Sig. (2-tailed) | 0.024 | 0.020 | 0.073 | 0.127 | 0.513 |
|  | N | 40 | 40 | 40 | 40 | 40 |
| Black male expectation of life at age 0 | Correlation Coefficient | .347* | .391* | 0.287 | .338* | 0.230 |
|  | Sig. (2-tailed) | 0.028 | 0.013 | 0.073 | 0.033 | 0.154 |
|  | N | 40 | 40 | 40 | 40 | 40 |
| Calculated life exp. Of White Male – Black Male | Correlation Coefficient | -0.164 | -0.178 | -0.103 | -0.231 | -0.251 |
|  | Sig. (2-tailed) | 0.311 | 0.272 | 0.526 | 0.152 | 0.118 |
|  | N | 40 | 40 | 40 | 40 | 40 |
| The composition percentage of White people in this state | Correlation Coefficient | -0.189 | 0.177 | -0.177 | .338* | 0.140 |
|  | Sig. (2-tailed) | 0.180 | 0.210 | 0.210 | 0.014 | 0.322 |
|  | N | 52 | 52 | 52 | 52 | 52 |
| The composition percentage of Black people in this state | Correlation Coefficient | -0.165 | -.308* | -0.106 | -.307* | -0.192 |
|  | Sig. (2-tailed) | 0.243 | 0.026 | 0.457 | 0.027 | 0.174 |
| | N | 52 | 52 | 52 | 52 | 52 |
| *2014 Poverty Rates (includes unrelated children)* | Correlation Coefficient | 0.067 | -0.111 | 0.083 | -0.026 | 0.026 |
| | Sig. (2-tailed) | 0.639 | 0.434 | 0.561 | 0.855 | 0.856 |
| | N | 52 | 52 | 52 | 52 | 52 |
| *Supplemental Poverty Measure (2010–2014 average)* | Correlation Coefficient | 0.204 | -0.004 | 0.225 | 0.003 | 0.048 |
| | Sig. (2-tailed) | 0.147 | 0.978 | 0.109 | 0.983 | 0.735 |
| | N | 52 | 52 | 52 | 52 | 52 |
| *High school graduate or higher* | Correlation Coefficient | 0.035 | 0.136 | -0.005 | 0.094 | -0.049 |
| | Sig. (2-tailed) | 0.804 | 0.335 | 0.972 | 0.508 | 0.730 |
| | N | 52 | 52 | 52 | 52 | 52 |
| *Bachelor's degree or higher* | Correlation Coefficient | .367** | .327* | .275* | 0.156 | -0.157 |
| | Sig. (2-tailed) | 0.008 | 0.018 | 0.048 | 0.270 | 0.266 |
| | N | 52 | 52 | 52 | 52 | 52 |
| *Advanced degree* | Correlation Coefficient | .444** | .440** | .366** | 0.184 | -0.069 |
| | Sig. (2-tailed) | 0.001 | 0.001 | 0.008 | 0.191 | 0.626 |
| | N | 52 | 52 | 52 | 52 | 52 |

**Correlation with Poverty Rate**

Looking at the poverty rates tables, the outcomes even apprised more prominently—the racist rates are positively correlated with "2014 Poverty Rates" and "Supplemental Poverty Measure" tables. In other words, there is a link that shows if the greater the number of Black people are suffering from racism content in a state, the more poverty rates are there in that state too. This rate has a significant correlation with P-value < 0.01 by the Spearman rho's correlation coefficient of 0.582 with "Supplemental Poverty Measure," and at P-value < 0.01 level by the coefficient of 0.460 with "2014 Poverty Rates" (Table 4).

*Table 4. Spearman's rho statistical analysis for percentage of Racist tweets compare to "Poverty Rates," "Educational Attainments," and "Percentage of Race Composition" in each state.*

| | | Number of Racist tweets Male | Number of Racist tweets Female | Racist tweets Male Before event | Racist tweets Female Before event | Racist tweets Male After event | Racist tweets Female After event | Racist rates (Racist / Sum) |
|---|---|---|---|---|---|---|---|---|
| *Composition of White people* | Correlation Coefficient | -.578** | 0.135 | -.578** | -.610** | 0.125 | -.603** | -.665** |
| | Sig. (2-tailed) | 0.000 | 0.341 | 0.000 | 0.000 | 0.376 | 0.000 | 0.000 |
| | N | 52 | 52 | 52 | 52 | 52 | 52 | 52 |
| *Composition of Black people* | Correlation Coefficient | .711** | -0.060 | .711** | .747** | -0.039 | .736** | .848** |
| | Sig. (2-tailed) | 0.000 | 0.673 | 0.000 | 0.000 | 0.785 | 0.000 | 0.000 |
| | N | 52 | 52 | 52 | 52 | 52 | 52 | 52 |
| *Composition of Native people* | Correlation Coefficient | -.316* | -0.008 | -.318* | -.325* | 0.022 | -.349* | -.423** |
| | Sig. (2-tailed) | 0.022 | 0.954 | 0.022 | 0.019 | 0.876 | 0.011 | 0.002 |
| | N | 52 | 52 | 52 | 52 | 52 | 52 | 52 |
| *2018 Poverty rate* | Correlation Coefficient | 0.233 | 0.028 | 0.232 | 0.270 | 0.073 | 0.242 | .304* |
| | Sig. (2-tailed) | 0.096 | 0.846 | 0.098 | 0.053 | 0.606 | 0.084 | 0.029 |
| | N | 52 | 52 | 52 | 52 | 52 | 52 | 52 |
| *2014 Poverty Rates* | Correlation Coefficient | .415** | 0.070 | .417** | .456** | 0.092 | .414** | .460** |
| | Sig. (2-tailed) | 0.002 | 0.623 | 0.002 | 0.001 | 0.518 | 0.002 | 0.001 |
| | N | 52 | 52 | 52 | 52 | 52 | 52 | 52 |
| *Supplemental Poverty Measure* | Correlation Coefficient | .561** | 0.054 | .566** | .602** | 0.066 | .578** | .582** |
| | Sig. (2-tailed) | 0.000 | 0.703 | 0.000 | 0.000 | 0.641 | 0.000 | 0.000 |
| | N | 52 | 52 | 52 | 52 | 52 | 52 | 52 |
| *High school graduate or higher* | Correlation Coefficient | -.530** | 0.042 | -.530** | -.556** | 0.015 | -.529** | -.573** |
| | Sig. (2-tailed) | 0.000 | 0.769 | 0.000 | 0.000 | 0.916 | 0.000 | 0.000 |
| | N | 52 | 52 | 52 | 52 | 52 | 52 | 52 |
| *Bachelor's degree or higher* | Correlation Coefficient | 0.085 | 0.137 | 0.080 | 0.065 | 0.100 | 0.099 | -0.086 |
| | Sig. (2-tailed) | 0.550 | 0.331 | 0.574 | 0.649 | 0.482 | 0.483 | 0.545 |
| | N | 52 | 52 | 52 | 52 | 52 | 52 | 52 |
| *Advanced degree* | Correlation Coefficient | 0.242 | 0.172 | 0.239 | 0.232 | 0.135 | 0.266 | 0.104 |
| | Sig. (2-tailed) | 0.084 | 0.224 | 0.087 | 0.099 | 0.340 | 0.057 | 0.463 |
| | N | 52 | 52 | 52 | 52 | 52 | 52 | 52 |

**Correlation with Educational Attainment**

The most impressive findings from the data are how educational attainment correlates with BLM and Racist groups. The more educated the people of an area are, the more users are there to support the BLM movement, and the less schooling level percentages areas had more racism rates.

Referring to Table 3, the rate of BLM tweets in each state positively correlates with States with higher "Bachelor's Degree" or "Advanced Degree" numbers. The related findings show a Spearman's Rho Correlation Coefficient of 0.444 between "Rate of all BLM tweets" and "Advanced Degrees" educational level with a p-value of $< 0.01$ for 52 states. Similarly, we have the Coefficient of 0.367 between the same factor and the "Bachelor's Degree or Higher" variable, attaining a p-value of 0.008.

On the other side of the coin, there is an adverse correlation between the number of Racists and "High School Graduate or Higher" statistics. It means the higher the rate of high school graduates in each state, there is lower racist content posted in that area (Table 4)

**Correlation with Race Composition**

Compared to the population compositions, the correlation with two Racist and Anti-BLM groups shows an outstanding sad novel finding. The rates of racist tweets are higher in areas with a higher Black population and are lower in states having supremacy of white people. The Anti-BLM tweets are more in states with more leading White compositions and lower in those with more prominent Black residents.

Regarding Table 4, the statistics for racist groups in correlation with population compositions show a strong relationship with the population of Black and White people.

## Discussion

The findings in analysis with these tables showed a positive connection between the racist tweets' rate and the life expectancy of Black males minus White males. Similar outcomes showed up for the rate of BLM tweets with Black's life expectations, implying a connection between the number of BLM tweets published in each area and this factor. It was slightly surprising that the rates of racist tweets extracted by our criteria – including offensive n-words – were noteworthily correlated with poverty measures of each area. Another surprising finding is how the educational attainment links to the rates of BLM and Racist tweets in each state.

There is a remarkable diversity in Whites' typical lives, whereas Blacks remain in the same economic class in a social environment. For instance, most white middle-class families live in average communities and send their children to typical schools. In contrast, many blacks in the same social class live in impoverished regions and foster their children at high-poverty schools [42]. Therefore, racial disparity, notably in poverty, persists in this modern age and influences many aspects of Black life [43]. The extent to which ethnic groups differ is combined regarding the BLM movement rates and racism in this study. Race and ethnicity study is complicated in this content as the United States of America has a vast, diverse population [44].

There was a direct connection between the upper percentages of bachelor's or advanced degrees and the support rates for the BLM hashtag regarding geography. On the other side, racism rates were also lower for states with more high school graduates. The population composition tables revealed disconsolate outcomes that Racist and Anti-BLM tweets directly correlate with the Black population. Therefore, it can show that Blacks are more prone to racism wherever they shaped a larger community.

Findings indicate that these cyber-discriminations have a real-world adverse link with the Black community's mattering lives.

## Why we studied "Black Lives Matter" with Respect to George Floyd?

In this study, we analyzed the #BlackLivesMatter movement and Racism rates in reference to the injustice killing of this Black man by a police officer. The tragedy of George Floyd's demise in 2020 was a justice failure that liberally impacted the number of related posts to anti-racism activity in the cyber world. He was an innocent Black man killed by a police officer on the street, coinciding with the murder of another Black man in Atlanta, which attracted a lot of attention. These two events proved to everyone that racism still exists in the United States and has not disappeared.

Battle against racism by #BlackLivesMatter started in late 2012, yet this event triggered it again. Racism is a hidden pain in our society, which is reflected again as a social media trend in this challenging year. We believe that science and research should solve or reflect real-world problems; that is why we considered this event as a turning point.

Now, this hashtag is a social movement to reflect endeavors, and we employed it in our research to present the highlights of these conversations compared to other statistical measures. The study's outcomes clearly show the changes regarding this event, while racist behaviors are still out there and a community is suffering.

The findings of this study considerably have a higher significance level when the results are focused on males. The primary outcome shows that the life expectancy of men, which is lower in default than Black women, is closely connected to the rates of racism content. Similarly, earlier surveys revealed that racial discrimination originates from physical health issues for Black people, yet the effects for males go further [45]. It is widely known that depression, stress, and suicide, over and above the medical complaints, are the underlying causes of being exposed to racism.

In a study [46], scientists found that a "sense of mastery" can help Black women reducing their stress, while the results were insignificant for men. Sense of mastery is a term referred to as the ability of self-control on challenging life issues, whereas mastering this ability on men could not reflect the same rates as women in overcoming emotional distress. In line with another study [47], scholars who sought to discover how educational attainment can defeat depressions concluded that women had conquered this battle better than men. They investigated 3.5k Black adults' records to uncover how superior education is bound up with lowering psychological distress. This study also showed that women are more kept from the harms of these depressive symptoms apart from men.

Overall, Black men manifest in more severe color separations that strike their health [48]. They are unjustifiably shot or killed by police beyond Black women [49], and Black youth experience more notable prejudiced treatment annually [50]. Social inequalities accelerated suicide, homicide, or sickness considerably more eminent than women's rates [51]. Finally, as a consequence of all these Black men's life expectancy at birth is far less than other groups such as white men, black women, or white women [52].

## Conclusions

This study aimed to determine how racism content toward Black people is correlated with life expectancy, poverty, and education. We shaped a criterion with the unfair death of a Black man – George Floyd – to mine inside the published tweets and extract them in a dataset with proper classification of supporting and opponent groups to this story. We labeled dataset records with their area name, and gender values were also added. One of the potential points of this research was the extensive number of tweets gathered to make the accuracy rates more eminent.

Blacks are susceptible to social inequalities, as this study showed the direct link between racism and their life expectancy. Previous research has shown how these disparities could lead to preterm birth, psychological issues, or self-destruction.

Overall, the most prominent finding to emerge from this study is that the conversation discussed on Twitter can be attributed as a sample speech of the offline world and directly influences Black people's lives when going nasty. This study showed that racism content is associated with the lower life expectancy of Black men. The analysis also showed a strong association in areas with immense racism content and the more poverty rates. We also found that higher education correlated with lower offensive language.

**Conflicts of Interest:** The authors declare no conflicts of interest.